\documentclass{elsart}

\usepackage{amssymb}
\usepackage{epsfig}
\begin{document}

\begin{frontmatter}



\title{Percolation transition of hydration water at hydrophilic surfaces}
\author{A. Oleinikova},
\ead{alla.oleinikova@heineken.chemie.uni-dortmund.de}
\author{I. Brovchenko and A. Geiger}
\address{Physical Chemistry, University Dortmund, Otto-Hahn-Str.6, D-44221}

\begin{abstract}
An analysis of water clustering is used to study the quasi-2D
percolation transition of water adsorbed at planar hydrophilic
surfaces. Above the critical temperature of the layering transition (quasi-2D liquid-vapor phase transition of adsorbed molecules) a percolation transition occurs at some threshold surface coverage, which increases with increasing temperature. The location of the percolation line is consistent with the existence of a percolation transition at the critical point. The percolation threshold at a planar surface is weakly sensitive to the size of the system when its lateral dimension increases from 80 to 150 $\mbox{\AA}$. The size distribution of the largest water cluster shows a specific two-peaks structure in a wide range of surface coverage : the lower- and higher-size peaks represent contributions from non-spanning and spanning clusters, respectively. The ratio of the average sizes of spanning and non-spanning largest clusters is about 1.8 for all studied planes.
 The two-peak structure becomes more pronounced with decreasing size of the planar surface and strongly enhances at spherical surfaces.  

\end{abstract}

\end{frontmatter}

\section{Introduction}
\label{Intro}
The existence of an infinite (spanning) network of hydrogen-bonded water molecules strongly affects the properties of aqueous systems and plays an important role in various technological and biological processes. In bulk liquid water such a three-dimensional network exists up to the liquid-vapor critical point \cite{Geiger,Geiger2,Guillot}. In the supercritical region a spanning water network appears via a percolation transition at some threshold value of the density, which increases with increasing temperature \cite{Kalinichev,Pal}. In aqueous solutions with rather hydrophilic solute, the formation of an infinite water network with increasing water content also occurs via a percolation transition \cite{perc1,Bates}. Whereas in solutions with rather hydrophobic solute, the formation of a spanning water network is preceded by the liquid-liquid phase separation \cite{perc2}. In an aqueous system with constituents, which are noticeably larger than water molecules, obviously, that spanning hydrogen-bonded network of water molecules should be formed rather via a 2D, than via a 3D percolation transition. Indeed, the percolation transition of water was found to be quasi-2D even in a solution of such a relatively small molecule, as tetrahydrofuran \cite{perc1}. The two-dimensional character of the water percolation transition in aqueous systems should gain more importance with increasing size of the solutes. 
\par 
Water at the surface of biomolecules (so-called hydration or biological water) strongly influences their structural and dynamical properties. In particular, the existence of a spanning network of hydration water in biosystems enables their biological functions \cite{Careri,pol}. The transformation from an ensemble of finite water clusters to an infinite water network with increasing hydration level occurs via a percolation transition, which was found to be two-dimensional in experiments and computer simulations of various biosystems (see \cite{lys1,lys2} and references therein). 
\par Despite the crucial role of the percolation of hydration water for the onset of biological activity, it was not studied yet even for the simplest model systems. The main goal of our paper is to study the (2D) percolation transition of water at smooth planar hydrophilic surfaces in order to create a basis for subsequent investigations of the percolation of hydration water in more complex, first of all biological, systems. Strictly speaking, such a transition is quasi-2D, since even at a smooth surface the adsorbed water molecules are not restricted to a single plane parallel to the adsorbate surface. Therefore, some deviations of the percolation transition of hydration water from conventional percolation in strict 2D systems can be expected. 
\par
Another goal of the present study is to locate the percolation threshold of the hydration water
relatively to the coexistence curve of the layering transition (quasi-2D condensation). It is expected, that the line of percolation transitions \cite{Kertesz} of so-called physical clusters \cite{Coniglio2,Vericat} should meet the thermodynamic phase transition at the critical point which is also a percolation point. This line was indeed observed for 2D and 3D Ising lattices \cite{Kertesz,Coniglio}, for the 3D lattice gas \cite{Campi} and for the Lennard-Jones fluid \cite{Campi2}. 
\par
In this paper we present the first computer simulation study of the quasi-2D percolation
transition of water molecules adsorbed at smooth planar hydrophilic surfaces. To locate the percolation transition, the water clustering with increasing surface coverage is analyzed above the critical temperature of the layering transition. Peculiarities of the water percolation at hydrophilic spherical surfaces are discussed.    
\section{Methods}
\label{Methods}
The percolation transition of adsorbed TIP4P water molecules \cite{TIP4P} was studied by constant-volume Monte Carlo (MC) simulations, using asymmetric slit-like pores, formed by a smooth hydrophilic wall and by a hard wall. The water-surface interaction with the hydrophilic wall was described by a (9-3) Lennard-Jones potential between the water oxygen and the wall with $\sigma$ = 2.5 $\mbox{\AA}$ and a well-depth {\it U$_0$} = -4.62 kcal/mol. The chosen potential is strong enough to provide a layering transition of water at this surface. The coexistence curve of this layering transition was obtained previously by MC simulations in the Gibbs ensemble \cite{water1}. 
\par 
The clustering of the water molecules was analyzed at several state points with {\it T} = 425 K , i.e. above the critical temperature of the layering transition at this surface ({\it T$_c$} $\approx$ 400 K \cite{water1}). Simulations were performed in cubic boxes with three different edge lengths: {\it L} = 80, 100 and
150 $\mbox{\AA}$. The surface coverage {\it C} =  {\it N}/{\it L$^2$} was varied by
putting various numbers {\it N} of water molecules in the simulation boxes. Periodic boundary conditions were applied in two directions parallel to the pore walls. Note, that the high localization of the water molecules in the vicinity of the hydrophilic wall was not sensitive to the variation of the width of the pore and edge length {\it L} of the box. 
\par
Water molecules were considered to belong to the same cluster if they are connected by a continuous path of
hydrogen-bonds (H-bond) \cite{Geiger,perc1,perc2,Stanley}. An H-bond between two water
molecules was assumed to exist, when the distance between the oxygen atoms  is $<$ 3.5 $\mbox{\AA}$
and the water-water interaction energy  is $<$ -2.4 kcal/mol. Various cluster properties were analyzed after every 1000th MC step and up to 1 million configurations were analyzed for each surface coverage \textit{C}. Each configuration was inspected to detect the possible presence of an "infinite" cluster, which spans the periodic simulation box at least in one
direction parallel to the hydrophilic wall. Then the probability to observe such a spanning cluster {\it R} was determined at each surface coverage. The occurence frequency of water clusters of various sizes {\it S} was described by the cluster size distribution {\it n$_S$}. The mean cluster size was calculated as {\it S$_{mean}$} = $\Sigma${\it n$_S$S$^{~2}$}/$\Sigma${\it n$_S$S}, where the largest cluster was excluded from the sums. {\it n$_S$S}/$\Sigma${\it n$_S$S} is the probability that a given water molecule is member of a finite cluster of size \textit{S}. 
\par 
We also analyzed some properties of the largest cluster (of size {\it S$_{max}$}). Firstly there is the size distribution {\it P(S$_{max}$)} of the largest cluster. At the percolation threshold the largest cluster should be a fractal object with a specific fractal dimension. The statistical
 self-similarity of infinite fractals leads to the following relationship between mass {\it m(r)} and linear size {\it r}:
\begin{eqnarray} 
\label{fract}
m(r) \sim r^{ d_f}  
\end{eqnarray}
where {\it d$_f$} is the fractal dimension. In our analysis of the fractal dimension of the largest water cluster, {\it m(r)} is the number of water molecules which belong to this cluster and which are located inside a sphere of
radius {\it r} around at a randomly chosen water molecule of the same cluster. We determined the distributions {\it m(r)} for each water molecule  of the largest cluster of the configuration
and averaged them for each surface coverage. The fractal dimension {\it d$_f$} was
determined from the fits of the data to equation (1) in the range {\it r} $<$ {\it L} for each
simulated system.         
\par  
 The percolation threshold, when a spanning cluster appears in an infinite system, can be considered to mark the lowest possible concentration, furnishing a full surface coverage . To locate the percolation threshold of water adsorbed on a hydrophilic surface, we used several criteria. Right at the percolation threshold the cluster size distribution {\it n$_S$} obeys the universal power law {\it n$_S$} $\sim$  {\it S$^{-\tau}$}, with exponents $\tau$ = 187/91 $\approx$ 2.05 \cite{Staufferbook} and $\tau$ $\approx$ 2.2 \cite{Jan} in the case of random 2D and 3D percolation, respectively. The mean cluster size {\it S$_{mean}$} diverges at the percolation threshold in an infinite system and passes through a maximum when
approaching the threshold in a finite system. The fractal dimension of the largest cluster at the percolation threshold is lower than the Euclidean dimension of the system and equal to {\it d$_f^{~2D}$} = 91/48 $\approx$ 1.896 and {\it d$_f^{~3D}$} $\approx$ 2.53 in the case of 2D and 3D percolation, respectively \cite{Staufferbook,Jan}. Note, that at the critical point of the 2D Ising model, which is a tricritical point for correlated percolation, the fractal dimension of the largest
cluster is about 1.95 (187/96) which is higher than \textit{d$_f$} = 187/91 for random percolation \cite{Stella}. However, we may neglect a possible trend toward tricriticallity in our simulations performed at 25 K above the critical temperature. 
\section{Results}
\label{Results}
\begin{figure} 
\begin{center}
\includegraphics [width=9cm]{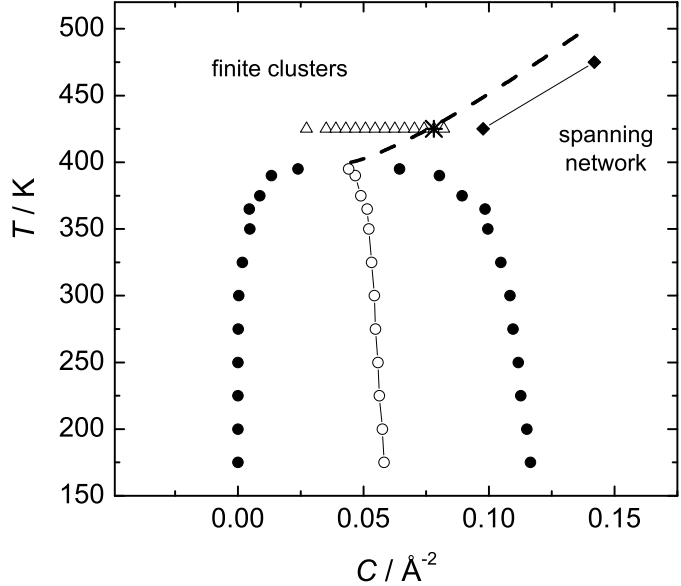}
\caption{Coexistence curve (solid circles) and diameter (open circles) of the layering transition of water \cite{water1} in terms of number of molecules per unit surface area {\it C}. The state points, where water clustering was studied at {\it T} = 425 K in the system with {\it L} = 80 $\mbox{\AA}$, are shown by open triangles. The percolation threshold and a possible percolation line are shown by the star and the dashed line, respectively. Percolation thresholds at the surface of a hydrophilic sphere of radius \textit{R$_{sp}$} = 15 $\mbox{\AA}$ are shown by solid diamonds. }  
\end{center}
\end{figure}
The coexistence curve of the quasi-2D layering transition of water near the studied hydrophilic surface, obtained by MC simulations in the Gibbs ensemble, is shown in Figure 1 for a slitlike pore of a width {\it H} = 24 $\mbox{\AA}$. Simulations of layering transitions in various pores with the same water-wall interaction show weak sensitivity of the coexistence curve to the pore shape and width (see Figure 11 in \cite{water1} and Figure 72 in Ref.\cite{handbook}). The shape of the coexistence curve of the layering transition of water in a wide temperature range corresponds to the 2D Ising model and the critical temperature is estimated as $\approx$ 400 K for the used water-wall interaction potential \cite{water1,handbook}. Extrapolation of the coexistence curve diameter (average surface coverage in the two coexisting phases) to the critical temperature gives the critical surface coverage of the layering transition {\it C$_c$} $\approx$ (0.045 $\pm$ 0.003) $\mbox{\AA}^{-2}$. To locate a percolation threshold, an analysis of the water clustering was performed at {\it T} = 425 K in the range of surface coverage {\it C} from 0.027 to 0.090 $\mbox{\AA}^{-2}$ (see Figure 1). From 175 to 1800 water molecules were placed in the simulation box. The state points, studied in the case of the smallest surface with {\it L} = 80 $\mbox{\AA}$, are shown by triangles in Figure 1. 
\begin{figure} 
\begin{center}
\includegraphics [width=9cm]{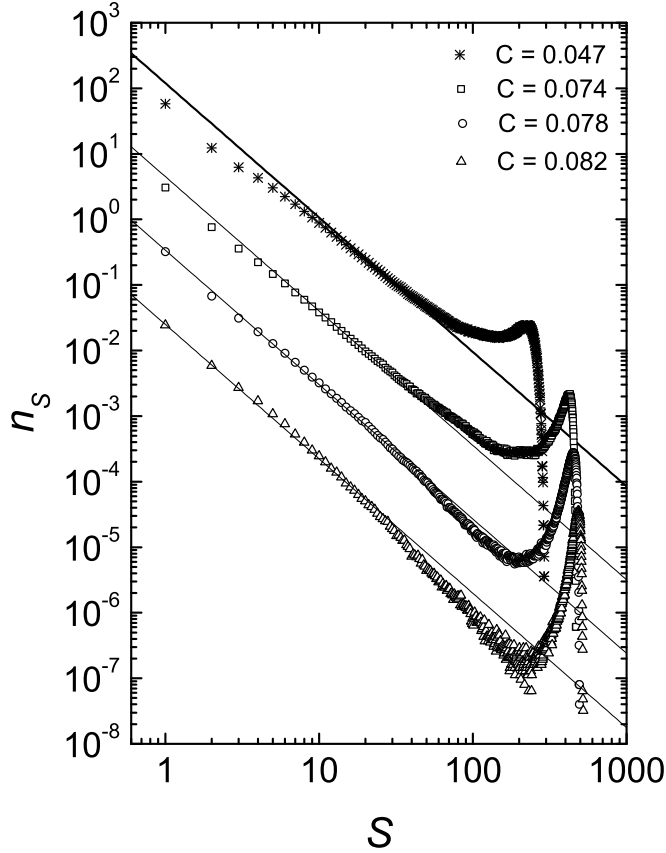}
\caption{Probability distribution {\it n$_S$} of clusters with {\it S} water molecules for several surface coverages {\it C} (in $\mbox{\AA}^{-2}$) below and above the percolation threshold {\it C$_0$} $\approx$ 0.078 $\mbox{\AA}^{-2}$ at the plane with {\it L} = 80 $\mbox{\AA}$). The critical power law {\it n$_S$} $\sim$ {\it S$^{ -2.05}$} is shown by the solid lines. The distributions are shifted vertically by one order of magnitude consecutively.}
\end{center}
\end{figure} 
\par
The percolation threshold can be located based on the analysis of the cluster size distributions {\it n$_S$}. At low surface coverage most of the water molecules belong to small clusters and {\it n$_S$} shows a rapid exponential decay with increasing {\it S}. Upon increasing the hydration level a hump appears in {\it n$_S$} at large {\it S} ({\it C} = 0.047 $\mbox{\AA}^{-2}$ in Figure 2). This hump reflects the truncation of the large clusters due to the finite size of the simulated system. At the percolation threshold the cluster size distribution {\it n$_S$} follows the power law behavior $\sim$ {\it S$^{ -\tau}$} in the widest range of cluster sizes with $\tau$ = 2.05 for 2D percolation. Figure 2 evidences that the percolation threshold of the adsorbed water at the plane with {\it L} = 80 $\mbox{\AA}$ occurs close to the surface coverage {\it C} = 0.078 $\mbox{\AA}^{-2}$ or
slightly below. When crossing the percolation threshold, deviations of {\it n$_S$} from the power law at large {\it S} before the hump, change the sign from positive to negative (compare {\it C} = 0.074 and 0.078 $\mbox{\AA}^{-2}$ in Figure 2). The negative deviations of {\it n$_S$} increase rapidly with increasing hydration above the percolation threshold ({\it C} = 0.082 $\mbox{\AA}^{-2}$).
\begin{figure} 
\begin{center}
\includegraphics [width=9cm]{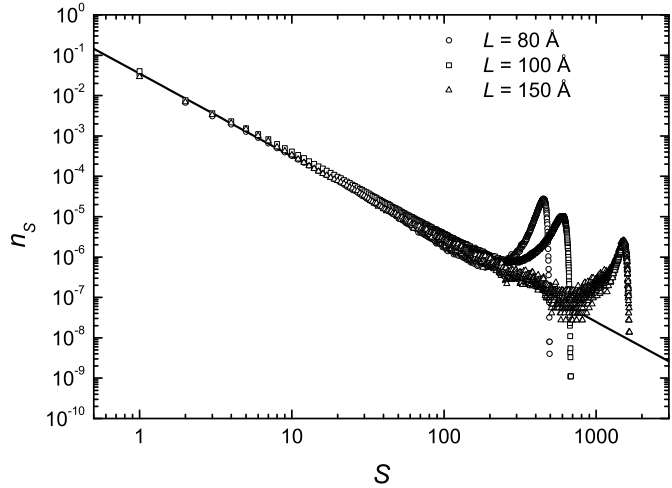}
\caption{Probability distribution {\it n$_S$} of clusters with {\it S} water molecules at planar surfaces of various sizes at surface coverages close to the percolation thresholds: {\it C} = 0.078 $\mbox{\AA}^{-2}$ (circles), 0.070 $\mbox{\AA}^{-2}$ (squares) and 0.078 $\mbox{\AA}^{-2}$ (triangles). The critical power law {\it n$_S$} $\sim$ {\it S$^{-2.05}$} is shown by a solid line.}
\end{center}
\end{figure}
\par A similar behavior of {\it n$_S$} is observed for the two other studied planar surfaces with {\it L} = 100 and 150 $\mbox{\AA}$. The percolation threshold is estimated at a coverage {\it C} close to 0.070 $\mbox{\AA}^{-2}$ for the plane with {\it L} = 100 $\mbox{\AA}$ and close to 0.078 for the plane with {\it L} = 150 $\mbox{\AA}$. The distributions {\it n$_S$} close to
the percolation threshold are compared for three studied planar surfaces in Figure 3. Note, that the variation of the surface coverage for the plane with {\it L} = 100 $\mbox{\AA}$ by 0.01 $\mbox{\AA}^{-2}$ near the coverage {\it C} = 0.070 $\mbox{\AA}^{-2}$ was rather coarse. Therefore, we conclude that the threshold surface coverage {\it C$_0$} obtained from the distributions {\it n$_S$}, is almost indistinguishable for the three studied surfaces and can be estimated as $\approx$ 0.078
$\mbox{\AA}^{-2}$.  
\begin{figure} 
\begin{center}
\includegraphics [width=9cm]{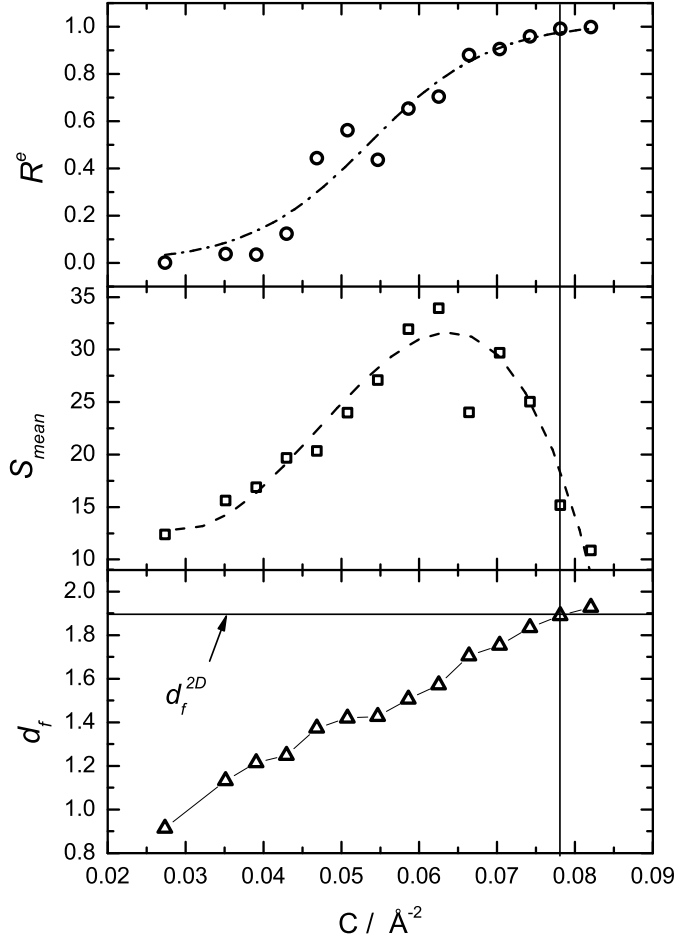}
\caption{2D percolation transition of water at the planar surface with {\it
  L} = 80 $\mbox{\AA}$. Spanning probability {\it R} (circles, upper
panel), mean cluster size {\it S$_{mean}$} (squares, middle panel) and fractal
dimension of the largest cluster {\it d$_f$}(triangles, lower panel) are shown
as function of water surface coverage {\it C}. The dot-dashed line is a fit of
{\it R} by the Boltzmann function. The dashed line is a guide for eyes only. The
vertical line indicates the threshold water coverages {\it C$_0$} estimated
from the behavior of {\it n$_S$}.} 
\end{center}
\end{figure}
\par
Various properties of water clusters at the planar surface with {\it L} = 80 $\mbox{\AA}$ are shown in Figure 4. The mean cluster size {\it S$_{mean}$} shows a shallow maximum at a surface coverage of about {\it C} $\approx$ 0.065 $\mbox{\AA}^{-2}$, i.e. below the percolation threshold {\it C$_0$}. The same behavior of {\it S$_{mean}$} is observed also for the other studied planar surfaces, in agreement with percolation theory. The fractal dimension of the largest cluster {\it d$_f$} is close to the value of the 2D percolation {\it d$_f^{2D}$} = 91/48 $\approx$ 1.896 at the threshold surface coverage {\it C$_0$} (see Figure 4, lower panel). This fact indicates the 2D character of the percolation transition of the adsorbed water. Values of {\it d$_f$} at various hydration levels for the three studied surfaces are compared in Figure 5 (lower panel). Evidently, {\it d$_f$} achieves the threshold value {\it d$_f^{2D}$} at about the same surface coverage {\it C$_0$}, independent from the size of the studied surface. This allows two conclusions: i) adsorbed water shows a 2D percolation threshold at {\it C$_0$} $\approx$ 0.078 $\mbox{\AA}^{-2}$ and ii) the cluster size distribution {\it n$_S$} and the fractal dimension of the largest cluster {\it d$_f$} at the percolation threshold are not very sensitive to the size of the simulated system.
\par 
\begin{figure} 
\begin{center}
\includegraphics [width=9cm]{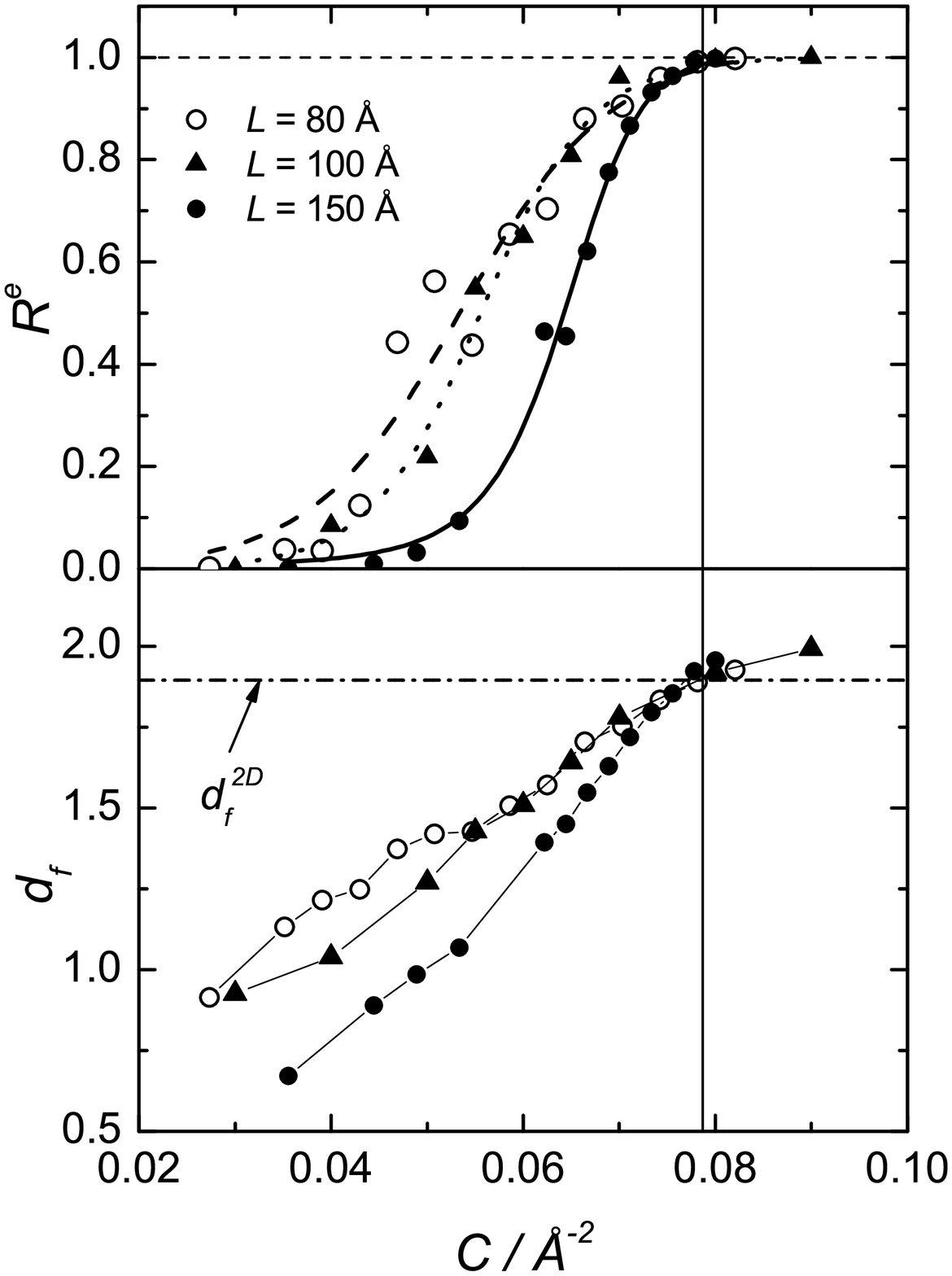}
\caption{Spanning probability  {\it R} (upper panel) and fractal dimension of
the largest water cluster {\it d$_f$} (lower panel) as functions of the surface
coverage {\it C} at the planar hydrophilic surfaces of various sizes. The fits of
{\it R} by the Boltzmann function are shown for the smallest, middle and largest
planar surfaces by dashed, dotted and solid lines, respectively. The threshold surface coverage {\it C$_0$} $\approx$ 0.078 $\mbox{\AA}^{-2}$ is shown by a vertical line.}
\end{center}
\end{figure}
The percolation threshold can also be located based on the spanning or wrapping
probability {\it R}, which is the probability to observe a spanning (percolating) cluster in a finite system of size {\it L}. This probability can be defined in various ways, which are called spanning rules. For instance, for 2D lattices there is {\it R$^1$}, the probability,
that a cluster spans in one direction \textit{only} (horizontally or vertically), {\it R$^e$}, 
the probability that the clusters spans {\it either} horizontally or vertically (or both), etc. In the present paper we used the spanning rule {\it R$^e$} to define a spanning probability.
 The dependency of the spanning probability on the occupancy variable (surface coverage) is shown in Figures 4 and 5 for different system sizes. The variation of {\it R} is steeper in larger systems. 
\par 
The dependency of the spanning probability {\it R} on the occupancy \textit{C} for various system sizes should cross at the percolation threshold. The value of {\it R} at this crossing point  depends on the spanning rule and on the dimensionality of the system, but should not depend on 
lattice structure, type of percolation (site or bond percolation), etc. \cite{Hovi}.
Figure 5 evidences that the {\it R$^e$(C)} for water adsorbed at surfaces of various size do not cross exactly in one point. The application of a sigmoidal (Boltzmann function) fit allows to locate the crossing point for the two smallest surfaces at about
{\it R$^e$} $\approx$ 0.7. However,  {\it R$^e$(C)} for the largest surface ({\it L} = 150 $\mbox{\AA}$) seems to be rather separated and meets {\it R$^e$(C)} for the smallest surface only when {\it R$^e$} $>$ 0.9. Note, that the value of the spanning probability at the threshold coverage  {\it C$_0$} $\approx$ 0.078 $\mbox{\AA}^{-2}$, obtained above from the behavior of {\it n$_S$} and {\it d$_f$}, exceeds 0.95 for all studied surfaces. 
\par 
\begin{figure} 
\begin{center}
\includegraphics [width=9cm]{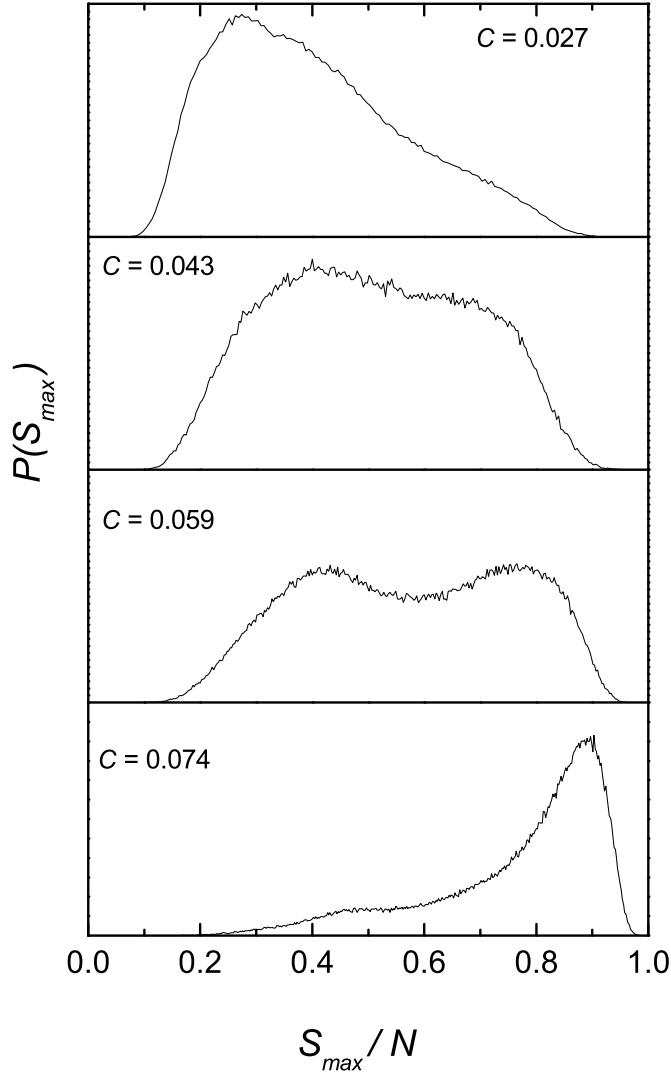}
\caption{Probability distribution {\it P(S$_{max}$)} of the size {\it S$_{max}$} of the largest
water cluster at the planar surface with {\it L} = 80 $\mbox{\AA}$ at various surface coverages {\it C} (in $\mbox{\AA}^{-2}$) normalized on the total number of water molecules {\it N}.}
\end{center}
\end{figure}
The probability distributions {\it P(S$_{max}$)} of the size {\it S$_{max}$} of the largest water cluster at the surface with {\it L} = 80 $\mbox{\AA}$ are shown in Figure 6 for several surface coverages {\it C}. At low hydration levels, {\it P(S$_{max}$)} has only a single maximum at low {\it S$_{max}$} (Figure 6, upper panel). With increasing surface coverage {\it P(S$_{max}$)}
shows a characteristic two-peak structure (middle panels) which remains noticeable close to the percolation threshold (lower panel). We have found, that the left peak is due to non-spanning largest clusters, whereas the right peak is due to spanning largest clusters. The two
contributions to {\it P(S$_{max}$)} are shown in Figure 7 for the planar surface with {\it L} = 100 $\mbox{\AA}$ at the surface coverage where  {\it R} is about 50 $\%$.    
\begin{figure} 
\begin{center}
\includegraphics [width=9cm]{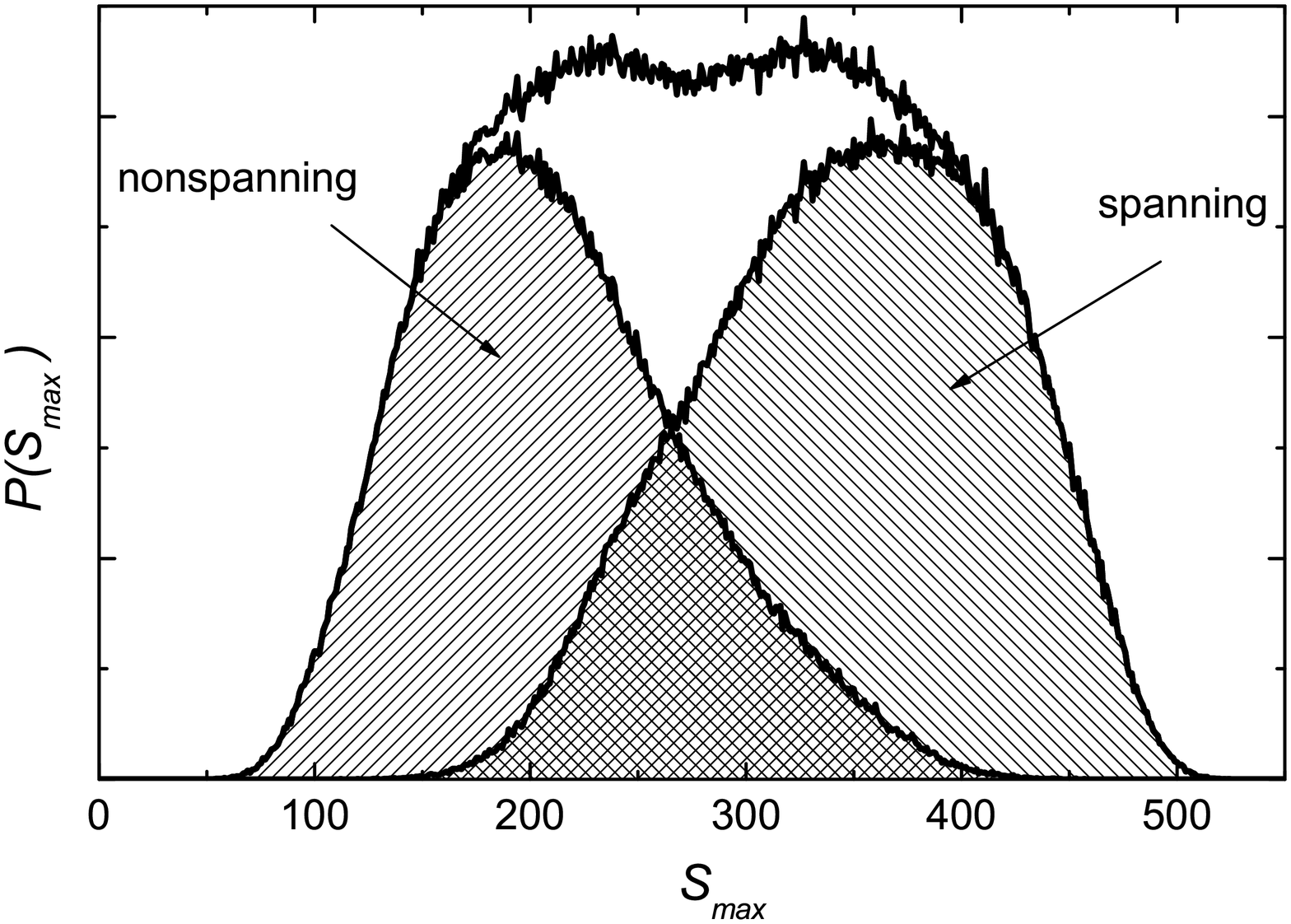}
\caption{Probability distribution {\it P(S$_{max}$)} of the size {\it S$_{max}$} of the largest
water cluster clusters at the planar surface with with {\it L} = 100 $\mbox{\AA}$ at the surface coverages {\it C} = 0.055 $\mbox{\AA}^{-2}$, where spanning and nonspanning largest clusters
exists with comparable probabilities. Size distributions of spanning and nonspanning largest clusters normalized on their respective probabilities, are shown as two dashed areas.     
}
\end{center}
\end{figure}
\begin{figure} 
\begin{center}
\includegraphics [width=10cm]{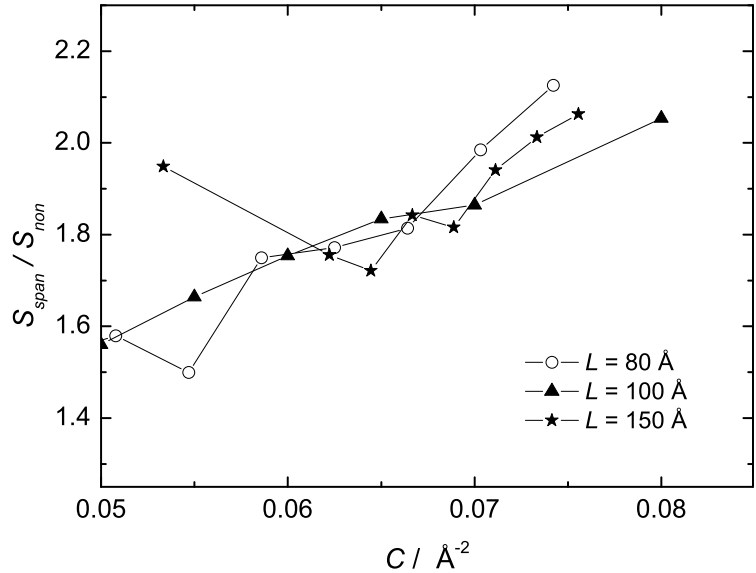}
\caption{Ratio of the average sizes of spanning {\it S$_{span}^{~av}$} and
nonspanning {\it S$_{non}^{~av}$} largest water clusters as a function of surface coverage.}
\end{center}
\end{figure}
\par We have calculated the average sizes of spanning {\it S$_{span}^{~av}$} and nonspanning {\it S$_{non}^{~av}$} largest clusters at various hydration levels. Their ratio shows no clear dependence on the system size and a weak tendency to increase with increasing surface coverage (see
Figure 8).  The ratio {\it S$_{span}^{~av}$}/{\it S$_{non}^{~av}$} can be well defined at the hydration level, where spanning and non-spanning largest clusters have comparable probabilities. In this range {\it S$_{span}^{~av}$}/{\it S$_{non}^{~av}$} is about 1.8 for all studied planes.
\begin{figure} 
\begin{center}
\includegraphics [width=9cm]{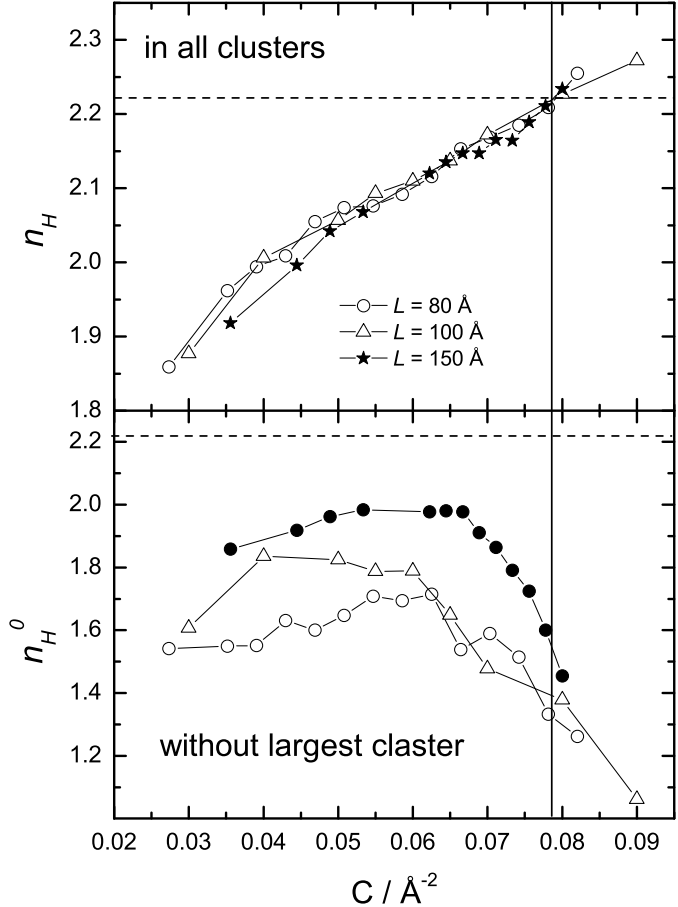}
\caption{Average number of hydrogen bonds, formed by each water molecule, as a function of surface coverage, calculated for all water molecules ({\it n$_H$}: upper panel) and for molecules, which do not belong to the largest cluster ({\it n$_H^{0}$}: lower panel).}
\end{center}
\end{figure}
\section{Discussion}
\label{Discussion}
Water adsorbed at smooth hydrophilic surfaces shows a well defined percolation transition at temperatures above the critical temperature of the layering transition. The fractal dimension of the largest cluster {\it d$_f$} at the percolation threshold, which is located by the power-law behavior of the cluster size distribution {\it n$_S$}, evidences the 2D character of the percolation transition. The surface coverage at the percolation threshold does not change noticeably with system size and we estimate it as {\it C$_0$} $\approx$ 0.078 $\mbox{\AA}^{-2}$ at {\it T} = 425 K. The spanning probabilities {\it R$^e$} for various system sizes intersect in one point at a surface coverage {\it C}, which is close to the percolation threshold value. 
\par 
Nevertheless, the behavior of the spanning probability {\it R} for the adsorbed water differs from that observed in conventional percolation studies of 2D lattices and continuous systems. For the definition of the spanning probability  {\it R$^e$} used in the present paper, values {\it R$^e$} at the intersection point were obtained from 0.69 \cite{Martins,Ziff2} to 0.81 \cite{Hovi} for 2D lattices and from 0.64 for percolation of hard and soft discs \cite{Hu} to $\sim$ 0.7 for 2D polymers \cite{Heermann}. 
The spanning probability at the intersection point in the case of adsorbed water ({\it R$^e$} $>$ 0.95) is noticeably higher than in strictly 2D systems. Note, that the
opposite trend should be expected due to the quasi-2D character of the simulated system:
{\it R} at the percolation threshold on 3D lattices is $\sim$
0.5 \cite{Martins}, i.e. lower than in 2D systems.
\par 
The obtained threshold water coverage is roughly about 70$\%$ of a water liquid monolayer, which is characterized by a surface coverage {\it C} $\approx$ 0.11 $\mbox{\AA}^{-2}$
at ambient temperature (see Figure 1). 
Taking into account, that for the studied systems the fraction of water molecules which stick out of the
first surface layer does not exceed 5$\%$  \cite{PRL}, we may conclude, that at
the percolation threshold water covers about 2/3 of the surface. It covers the whole surface more or less "homogeneously" but by a ramified fractal, which needs only 2/3 of the material of monolayer. 
This threshold surface
coverage can be compared with the threshold value of the occupancy variable in
2D lattice models. The honeycomb lattice with 3 neighbors and the square lattice
with 4 neighbors seem to be the most relevant to water adsorbed at a
hydrophilic surface \cite{handbook,BrovGeiger}. For these lattices the
threshold values of the occupancy variable are $\approx$ 0.70 and $\approx$ 0.59 for site
percolation and $\approx$ 0.65 and 0.5 for bond percolation, respectively \cite{Staufferbook}. So, quasi-2D correlated site-bond percolation of adsorbed water occurs at a 
threshold occupancy, which is rather close to the thresholds for random site and bond percolation in 2D lattices. On the other hand, water coverage at the threshold is smaller
than the corresponding value ($\approx$ 0.85 \cite{Hu}) for the random percolation of hard (non-overlapping) discs at the surface. Note, that the occupancy variables discussed above refer to closed packing of discs and should not be mixed
with so-called space occupation probability, which is $\approx$ 0.45 at the threshold for 2D site
percolation in various lattices \cite{Zallen} (see paper \cite{lys1} for more details). So, we have found a strong similaritity of the percolation transition of the adsorbed quasi-2D water with conventional percolation in 2D lattices. 
Note, that site-bond percolation transition in bulk liquid water was found to be similar to the percolation transition in 3D diamond lattice  \cite{Geiger,Geiger2}. 
\begin{figure} 
\begin{center}
\includegraphics [width=9cm]{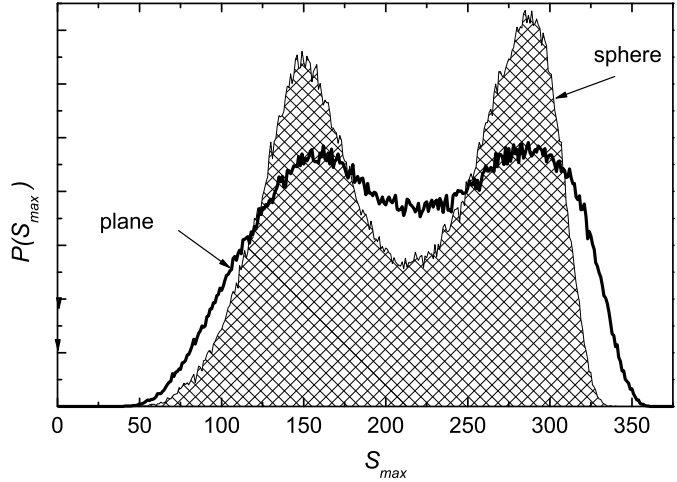}
\caption{Probability distribution {\it P(S$_{max}$)} of the size {\it S$_{max}$} of the largest
water cluster at the planar surface
  with  {\it L} = 80 $\mbox{\AA}$ and \textit{N}= 350, compared with {\it
    P(S$_{max}$)} at the surface of a 
hydrophilic sphere of radius  {\it R$_{sp}$} = 15 $\mbox{\AA}$ and \textit{N}= 375 (dashed area). 
}
\end{center}
\end{figure}
\begin{figure} 
\begin{center}
\includegraphics [width=9cm]{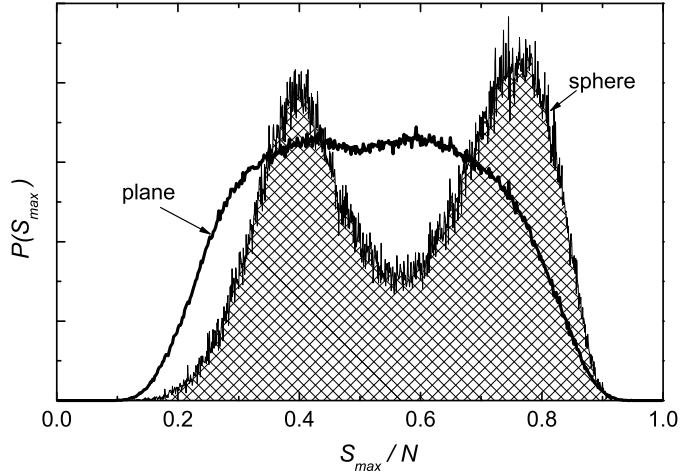}
\caption{Probability distribution {\it P(S$_{max}$)} of the size {\it S$_{max}$} of the largest
water cluster at the planar surface
  with  {\it L} = 100 $\mbox{\AA}$ (line, shown also in Figure 7) is compared with {\it P(S$_{max}$)} at the surface of a 
hydrophilic sphere of radius  {\it R$_{sp}$} = 30 $\mbox{\AA}$ at {\it C} =
0.088 $\mbox{\AA}^{-2}$ (dashed area). 
}
\end{center}
\end{figure}
\par To examine the possible location of the line of water percolation transitions relatively to the coexistence region of the layering transition we used our studies of the percolation threshold at the surface of a sphere \cite{lys1,PRL}. The location of the water percolation threshold at {\it T} = 425 and 475 K at the surface of a sphere with radius {\it R$_{sp}$} = 15 $\mbox{\AA}$ is shown in Figure 1 (solid diamonds). The shift of the percolation threshold at the spherical surface toward higher values of surface coverage is due to an increasing number of water molecules in the second hydration shell (see Ref. \cite{PRL}). The opposite trend is expected for the percolation transition on the inner surface of a cylinder, where indeed the layering transition is found slightly shifted towards lower surface coverages \cite{handbook}. Taking into account, that the shift of the percolation threshold of surface water with temperature is rather universal for various surfaces \cite{PRL}, we may schematically draw the line of the percolation transitions of water at the planar surface, as it is shown in Figure 1. Obviously, our results are consistent with the theoretical expectations \cite{Kertesz,Coniglio2} and with the results of
simulations of the lattice gas and LJ fluids, which show that the line of percolation transitions meets the coexistence curve at the critical point \cite{Campi,Campi2}. 
\par At subcritical temperatures, the percolation transition can be
prevented by phase separation, which appears as the formation of droplets of the
second phase. In computer simulations, however, the formation of droplets is
hampered by the finite size of the simulated systems and the second phase can be
detected only deeply inside the two-phase region \cite{Binderdrop}.
\par   
The relative insensitivity of the percolation threshold of hydration water to the size of the studied
system is very helpful, since it allows meaningful studies of percolation in
rather small systems. Simulations with just a few hundred water molecules are enough to accurately locate a 2D percolation threshold. This fact, which is especially useful for the study of networks
of the surface water in biological systems, obviously originates from the invariance of
water clustering and percolation in terms of the number of water-water hydrogen
bonds \cite{PRL}. 
\par Figure 9 shows the average number of water-water H-bonds in the three
studied systems at various hydration levels, when averaging was performed over
all clusters ({\it n$_H$}, upper panel) and when the largest cluster was
excluded from consideration ({\it n$_H^0$}, lower panel). Evidently, the
finite-size effect is noticeable for non-largest clusters only. The percolation
threshold for all planes is located at {\it n$_H$} = 2.22 and this value varies no more
on $\pm$ 0.1 for other temperatures and systems, including biosystems
\cite{PRL}. 
\par  
Finally, we would like to discuss the two-peak structure of the size distribution
of the largest cluster {\it P(S$_{max}$)} shown in Figures 6 and 7. This structure was observed
recently for 2D lattice models \cite{Sen}. The ratio {\it
  S$_{span}^{~av}$}/{\it S$_{non}^{~av}$} for lattices was found between 1.6 and 1.7 with
some week dependence on dimensionality. We observed a slightly higher value
of this ratio of about 1.8, which indicates a larger difference between the sizes of the spanning and
nonspanning largest clusters in the continuous system. With increasing system size
the two-peak structure of {\it P(S$_{max}$)} becomes less and less
pronounced. As the ratio {\it S$_{span}^{~av}$}/{\it S$_{non}^{~av}$} does not
depend on the size of simulated plane (see Figure 8), this effect should be
attributed to the increase of the width of the distributions of spanning
  and non-spanning largest clusters. 
\par The two-peak structure of {\it P(S$_{max}$)} is strongly enhanced at
spherical surfaces \cite{lys1,lys2}. In Figure 10, we compare {\it P(S$_{max}$)} for the
plane with {\it L} = 80 $\mbox{\AA}$ and for the spherical surface of radius {\it
  R$_{sp}$} = 15 $\mbox{\AA}$ with approximately the same number of adsorbed
water molecules ({\it N} = 350 and 375,
respectively). Evidently, the distributions of the spanning and nonspanning
largest clusters are much narrower at the spherical surface. Besides, the ratio
 {\it S$_{span}^{~av}$}/{\it S$_{non}^{~av}$} is about 2 in the latter
 case. Both factors can be responsible for the enhancement of the two-peak
 structure of {\it P(S$_{max}$)} at the spherical surface. 
\par One peculiarity of the spherical surface reported in reference
\cite{lys2} is the independence (or weak dependence) of the two-peak structure of {\it
  P(S$_{max}$)} on the surface size. In Figure 11 we compare {\it
  P(S$_{max}$)} for the planar and for the spherical surface with comparable surface area $\sim$ 10000
$\mbox{\AA}^2$ at the surface coverages, where the spanning and non-spanning
configurations have roughly equal probabilities. Comparison of Figures 10 and 11 shows, that the two-peak structure of {\it P(S$_{max}$)}  diminishes with increasing size of the plane, whereas this is not the case for spheres. The ratio
 {\it S$_{span}^{~av}$}/{\it S$_{non}^{~av}$} was also found insensitive to
  the size of the sphere and remains about 2 for spheres of
radius 15, 30 and 50 $\mbox{\AA}$ \cite{lys1}. This phenomenon indeed deserves further studies. The enhancement of the two-peak
  structure at spherical surfaces has an important consequence: the
spanning and non-spanning clusters can be distinguished even in cases,
where a spanning cluster can not be defined in the conventional way. The
  pronounced two-peak structure of {\it P(S$_{max}$)} allows the separation of
  spanning and non-spanning clusters at the surface of other finite objects,
  for example, at the surface of biomolecules \cite{lys1,lys2}.    



\end{document}